\documentclass[10pt]{article}          
\usepackage{mathrsfs,amsfonts}
minus.4em \belowdisplayshortskip=2mm plus.2em minus.4em

\makeatletter 
\@addtoreset{equation}{section}  \makeatother

\title{\centerline{\normalsize }
{\bf Generalized Solutions for Quantum Mechanical Oscillator on K\"{a}hler
  Conifold}}

\author{ {\bf Pulak Ranjan Giri\thanks{e-mail :pulakranjan.giri@saha.ac.in}}\\
\normalsize Saha Institute of Nuclear Physics, 1/AF Bidhan-Nagar,  Calcutta
700064, India}

\date{\today}

\begin{document}\maketitle 
\begin{abstract} \noindent\small
We study the possible generalized boundary conditions and the corresponding
solutions for the quantum mechanical oscillator model on K\"{a}hler
conifold. We perform it by self-adjoint extension of the the initial domain of
the effective radial Hamiltonian. Remarkable effect of this generalized
boundary condition is that at certain boundary condition the orbital angular
momentum degeneracy is restored! We also recover the known spectrum in our
formulation, which of course correspond to some other boundary condition.\\

\end{abstract} 
\section{\small{\bf {Introduction}}} \label{in} 
Quantum oscillator is an important model in various branches of physics, i.e,
quantum mechanics, quantum field theory, string theory and gravity due to its
exact solvability and overcomplete symmetry. The symmetry  is manifested
through angular momentum degeneracy of the energy spectrum. It is also
possible to separate the differential equation with respect to variables in
few coordinate systems. The overcomplete symmetry let the harmonic oscillator
to remain exactly solvable even after  some deformation of the potential is
made. So the symmetry is the prime issue which gives the the harmonic
oscillator such a status in different fields of study. But in quantum
mechanical oscillator on  K\"{a}hler conifold \cite{kahler}, proposed in
Ref. \cite{bellucci}, on the other hand this symmetry is generally broken. It
is an important model nevertheless, because it is solvable and it is defined
on  K\"{a}hler conifold, which is  a curved space. In string theory and
gravity  K\"{a}hler space \cite{zumino, bowick}  gets immense importance.  It
is a four dimensional quantum oscillator on the $(\nu,\epsilon)$ parametric
family of  K\"{a}hler conifolds related to the complex projective space $CP^2$
for $\nu =1$ and $\epsilon =1$ and four dimensional Lobacewski space $\mathcal
L_2$ for $\nu =1$ and $\epsilon =-1$.

Now the question is whether it is possible to retain the degeneracy of angular
momentum in the energy spectrum of the oscillator defined in
Ref. \cite{bellucci}. The answer is yes! In our present work we are going to
address this issue. We will basically perform an one parameter family of
self-adjoint extension \cite{reed} of the initial domain of the radial
Hamiltonian of the harmonic oscillator \cite{bellucci} by von Neumann method
\cite{reed}. This will help us  to construct  a generalized boundary condition.
We will show that for a  particular value of the extension parameter we can in
fact recover the angular momentum degeneracy in the energy spectrum. Not only
that, it is also possible to get the previously obtained result
\cite{bellucci} for another value of the parameter and other results.
 
However, the importance of self-adjointness of an operator is far
fundamental. As we know evolution of a quantum system is dictated by unitary
group and the generator of this group is the Hamiltonian itself. According to
Stone's theorem \cite{reed} generators of unitary group (in this case
Hamiltonian) should be self-adjoint. So for a non self-adjoint operator we
should search for a self-adjoint extension if possible. If the system has many
self-adjoint extensions then different self-adjoint extensions should unveil
different physics for the system.

The paper is organized as follows: In Sec.~\ref{os}, we discuss about the
quantum mechanical oscillator on K\"{a}hler conifold. In Sec.~\ref{ra}, we
perform the self-adjoint  extension of the radial Hamiltonian and we make some
observations for some particular value of the extension parameter
$\omega_0$. Here we show that it is actually possible to retain degeneracy in
the energy spectrum (symmetry of the system).  We discuss in Sec.~\ref{con}.

\section{\small{\bf {Quantum mechanical oscillator on K\"{a}hler conifold }}} 
\label{os} 

The Hamiltonian for the system is given by
\begin{eqnarray}
{\widehat{\cal H}}= -\hbar^2 g^{a \bar b}\partial_a\partial_{\bar b} +V_{osc},
\label{ham}
\end{eqnarray}
where the metric is of the form
\begin{eqnarray}
 g_{a\bar b}= \frac{\nu r_0^2 (z\bar z)^{\nu -1}}{2(1+\epsilon (z\bar z)^\nu
)} \left( \delta_{a b}-\frac{1-\nu+ \epsilon (z\bar z)^\nu}{z\bar z\;
(1+\epsilon (z\bar z)^\nu)}\bar z^a z^b\right),
\label{metric}
\end{eqnarray}
and the oscillator potential is given by
\begin{eqnarray}
V_{osc}= \omega^2 g^{\bar a b}\partial_{\bar a}K \partial_b K= \frac{\omega^2
r_0^2}{2} (z\bar z)^\nu.
\label{pot}
\end{eqnarray}
$K$  the potential of the  Ka\"{a}hler structure is given by
\begin{eqnarray}
K=\frac{r^2_0}{2\epsilon} \log (1+\epsilon(z\bar z)^\nu),\quad \nu >0; \quad
\epsilon=\pm 1,
\label{kpot}
\end{eqnarray}
In order to investigate the maximum possible solutions of the problem we need
to consider the eigenvalue problem, which is
\begin{eqnarray}
{\widehat{\cal H}}\Psi= E\Psi,
\label{eigen}
\end{eqnarray}
Equation  (\ref{eigen})  can be separated out in spherical coordinates
\begin{eqnarray}
z^1& =&
r^{\frac{1}{\nu}}\cos\frac{\beta}{2}\exp\left[\frac{i}{2}(\alpha+\gamma)\right],\nonumber\\
z^2& =& - ir^{\frac{1}{\nu}}\sin\frac{\beta}{2}\exp\left[-\frac{i}{2}(\alpha
-\gamma)\right],
\label{spherical}
\end{eqnarray}
if we consider the trial wavefunction of the form
\begin{eqnarray}
\Psi=\psi(r) D^j_{m,s}(\alpha,\beta,\gamma).
\label{wavef}
\end{eqnarray}
Here $\alpha\in [0,2\pi)$, $\beta\in [0,\pi]$ and $\gamma\in [0,4\pi)$, and
$r$ is a dimensionless radial coordinate taking values in the interval  $
[0,\infty)$  for $\epsilon=+1$, and in $[0, 1]$ for $\epsilon=-1$.  In the
Wigner function $ D^j_{m,s}(\alpha,\beta,\gamma)$ $j$, $m$ denote orbital and
azimuthal quantum numbers and corresponding operators are $\widehat
J^2,\widehat J_3$ respectively, while $s$ is the eigenvalue of the operator
$\widehat J_0$.
\begin{eqnarray}
&&\widehat{J}_0\Psi=s\Psi,\label{J0}\\ &&\widehat{\bf J}^2\Psi
=j(j+1)\Psi,\quad \widehat{J}_3\Psi=m\Psi ,\\ && m,s=-j,-j+1,\ldots , j-1,
j\;\; \mbox{where}~~ j=0,1/2,1,\ldots \label{mj}
\end{eqnarray}
The  volume element reads
\begin{eqnarray}
dV_{(4)}=\frac{\nu^2r_0^4}{32}\frac{r^3}{(1+ \epsilon r^2)^3}\sin\beta
drd\alpha d\beta d\gamma. \label{volume}
\end{eqnarray}
Separating the differential equation we get the radial eigenvalue equation  of
the form
\begin{eqnarray}
H(r)\psi(r)= E\psi(r),
\label{radialeigen}
\end{eqnarray}
where
\begin{eqnarray}\textstyle{
H(r)=-\frac{\hbar^2(1+\epsilon r^2)^2}{2r_0^2}\left[\frac{d^2}{dr^2} +
\frac{3+\epsilon r^2}{1+\epsilon r^2}\frac{1}{r} \frac{d}{dr} +
\frac{\epsilon\omega^2 r_0^4}{\hbar^2(1+ \epsilon r^2)^2} -\frac{\epsilon
\delta^2}{1+\epsilon r^2} -\frac{4\nu j(j+1)+4(1-\nu)s^2}{\nu^2 r^2(1+\epsilon
r^2)}\right]}
\label{radialham}
\end{eqnarray}
%
%
We now move to the next section to discuss the self-adjointness of the radial
Hamiltonian $H(r)$ of  Eq.(\ref{radialham}).

\section{\small{\bf {Self-adjointness of the radial Hamiltonian }}} \label{ra} 
The effective radial Hamiltonian $H(r)$, Eq.(\ref{radialham}) is formally
self-adjoint, but formal self-adjointness does not mean that it is
self-adjoint on a  given domain \cite{dunford}. This operator $H(r)$ belongs
to unbounded differential operator defined on a Hilbert space. As we have
mentioned in our introduction that we will perform self-adjoint extension of
the operator $H(r)$ by von Neumann's method \cite{reed}, so for the shake of
completeness here  we briefly review the von Neumann method.

Let us consider an unbounded differential operator $T$ defined over a Hilbert
space $\mathcal H$ and consider a domain  $D(T)\subset \mathcal H$ for the
operator $T$ such that it becomes symmetric on the domain  $D(T)\subset
\mathcal H$. Note that the operator $T$ is called symmetric or Hermitian if
$(T\phi,\chi)= (\phi, T\chi)~~ \forall \phi,\chi \in D(T)$, where (.~,~.) is
the inner product defined over the Hilbert space $\mathcal H$. Let
$D(T^\dagger)$ is the domain of the corresponding  adjoint operator
$T^\dagger$. The operator $T$ is self-adjoint iff $T= T^\dagger$ and $D(T)=
D(T^\dagger)$.

We now state the criteria of self-adjointness of  a symmetric operator $T$
according to von Neumann method. We need to find out the the deficiency
subspaces (it is actually a null space) $D^\pm \equiv \mbox{Ker}(i\mp
T\dagger)$ and the deficiency indices $n^\pm(T) \equiv \dim(D^\pm)$.
Depending upon $n^\pm$,  $T$ is classified as \cite{reed}:
\begin{list}{\arabic{enumi})}{\usecounter{enumi}}

\item $T$ is essentially  self-adjoint if $n^+= n^- = 0$.

\item $T$ has a $n$-parameter family of self-adjoint extension if $n^+ = n^-=
n \ne 0$.

\item $T$ has no self-adjoint extension if $n^+\ne n^-$. In this case $T$ is
called maximally symmetric.
\end{list}

We now return to the discussion of our effective radial differential operator
$H(r)$. This operator  is symmetric in the domain
\begin{eqnarray}
D(H(r)) = \{\phi(r): \parbox[t]{9cm}{\mbox{$\phi(r) = \phi'(r) = 0 $},
  absolutely continuous, square integrable  over the full range  with  measure
  \mbox{$d\mu$} \}\,.}
\label{domain1}
\end{eqnarray}
where $d\mu = \frac{r^3}{(1+ \epsilon r^2)^3}dr$, $\phi'(r)$ is the derivative
of $\phi(r)$ with respect to $r$.  The domain of the adjoint operator
$H^\dagger(r)$, whose differential expression is same as $H(r)$ due to formal
self-adjointness, is given by
\begin{eqnarray}
D^\dagger(H(r)) = \{\phi(r): \parbox[t]{9cm}{ absolutely continuous, square
    integrable  over the full range  with  measure \mbox{$d\mu$} \}\,,}
\label{domain2}
\end{eqnarray}
 $H(r)$ is obviously not self-adjoint \cite{reed}, because
\begin{eqnarray}
D(H(r))\ne D(H^\dagger(r))
\label{nonselfad}
\end{eqnarray}
So we may ask whether there is any possible self-adjoint extension \cite{reed}
for the problem? To answer this question we need to investigate whether there
is any square-integrable solution for the differential equations
\begin{eqnarray}
H(r)^\dagger \phi^\pm = \pm i\phi^\pm
\label{imaginarysol}
\end{eqnarray}

Eq. (\ref{imaginarysol})  can be transformed into Hypergeometric \cite{abr}
differential equation  upon transformation
\begin{eqnarray}
r= \cases{ \tan\theta,&  for $\epsilon=1$; \cr \tanh\theta, & for $\epsilon= -
 1$;}
\label{trans}
\end{eqnarray}
and taking the trial solution of the form
\begin{eqnarray}
\phi^\pm= \cases{ \sin^{j_1-1}\theta\cos^{\delta}\theta\; \psi^\pm, & for
$\epsilon=1$; \cr  \sinh^{j_1-1}\theta\cosh^{-\delta - 2a^\pm}\theta\;
\psi^\pm, & for $\epsilon=-1$.}
\label{cons1}
\end{eqnarray}

The transformed differential equation is given by
\begin{eqnarray}
t(1-t)\frac{d^2\psi^\pm}{dt^2} + \left[c-(a^\pm +b^\pm +1)t
\right]\frac{d\psi^\pm}{dt} - a^\pm b^\pm\psi^\pm,
\label{hyper}
\end{eqnarray}
where
\begin{eqnarray}
a^\pm = \frac{1}{2}\left(1+j_1+ \epsilon\delta - \sqrt{ \frac{\pm 2 r_0^2
i}{\epsilon\hbar^2} +4+ \frac{\omega^2 r_0^4}{\epsilon^2\hbar^2}}\right),
b^\pm = \cases{- a^\pm +\delta+j_1+1, & for $\epsilon=1$; \cr a^\pm +\delta,
& for $\epsilon= -1$;}\nonumber
\end{eqnarray}
\begin{eqnarray}
c= j_1+1,\nonumber j_1^2 = \frac{4j(j+1)}{\nu} +1-\frac{4(\nu-1)s^2}{\nu^2},
\delta^2= \frac{4s^2}{\nu^2} + \frac{\omega^2 r_0^4}{\hbar^2},\nonumber
\end{eqnarray}
\begin{eqnarray}
t= \cases{\sin^2\theta,  & for $\epsilon=1$; \cr \tanh^2\theta   & for
$\epsilon= -1$.}
\label{}
\end{eqnarray}
The squareintegrable solutions  of the deficiency space, apart from
normalization  is given by
\begin{eqnarray}
\phi^\pm=\cases{ D t^{\frac{c -2}{2}}(1-t)^{\frac{b^\pm + a^\pm -c}{2}}\;
_2F_1(a^\pm, b^\pm; c; t),\; & for $\epsilon=1$; \cr D
(\frac{t}{1-t})^{\frac{c -2}{2}}(\frac{1}{1-t})^{-\delta -2a^\pm}\;
_2F_1(a^\pm, b^\pm, c; t) & for $\epsilon= - 1$,}
\label{constant}\end{eqnarray}
where $_2F_1$ is the Hypergeometric function \cite{abr}.

The existence of these complex eigenvalues  of $H(r)^\dagger$  signifies that
$H(r)$ is not self-adjoint. The solution $\phi^\pm$ belong to the null space
$D^\pm$ of $H(r)^\dagger \mp i$. where $D^\pm \in D^\dagger(H)$. The dimension
of $D^\pm$ are known as deficiency indices $n^\pm$ and is defined by
\begin{eqnarray}
n^\pm =  \dim(D^\pm)
\label{deficiencyindices}
\end{eqnarray}
%



Since in our case the deficiency indices $n^+ = n^- = 1$, we can have a
1-parameter family of self-adjoint extension of $H(r)$. The selfadjoint
extension of $H(r)$ is given by $H(r)^{\omega_0}$ with domain
$D(H(r)^{\omega_0})$, where
\begin{eqnarray}
D(H(r)^{\omega_0})= \{ \psi(r)= \phi(r)+ \phi^+(r) + e^{i\omega_0}\phi^-(r) :
    \phi(r)\in D(H(r)), \omega_0\in \mathbb{R} (\bmod 2\pi)\}.
\label{selfdomain}
\end{eqnarray}
The bound state  solution of $H(r)^\omega$ is given by
\begin{eqnarray}
\psi(r) =\cases{ C t^{\frac{c -2}{2}}(1-t)^{\frac{b  + a  -c}{2}} \; _2F_1(a,
b; c; t), & for  $\epsilon=1$; \cr C (\frac{t}{1-t})^{\frac{c
-2}{2}}(\frac{1}{1-t})^{-\delta - 2a} \; _2F_1(a, b, c; t), & for $\epsilon= -
1$;}
\label{beigenvalue}\end{eqnarray}
where
\begin{eqnarray}
a = \frac{1}{2}\left(1+j_1+ \epsilon\delta - \sqrt{ \frac{2 r_0^2
E}{\epsilon\hbar^2} +4+ \frac{\omega^2 r_0^4}{\epsilon^2\hbar^2}}\right)\,, b
= \cases{- a  +\delta+j_1+1, & for $\epsilon=1$; \cr a+\delta   & for
$\epsilon= -1$;}\nonumber
\end{eqnarray}
\begin{eqnarray}
c= j_1+1, t= \cases{\sin^2\theta,   & for $\epsilon=1$; \cr  \tanh^2\theta, &
for $\epsilon= -1$;}
\end{eqnarray}
and $C$ is the normalization constant.  To find out the eigenvalue we  have to
match the function $\psi(r)$ with the domain  (\ref{selfdomain}) at
$r\rightarrow 0$. In the limit  $r\rightarrow 0$,
\begin{eqnarray}
\psi(r) \to \cases{C t^{\frac{c -2}{2}}(1-t)^{\frac{b  + a  -c}{2}}
\left[\Gamma_1 + (1-t)^{c-a-b}\Gamma_2 \right], & for  $\epsilon=1$; \cr C
t^{\frac{c -2}{2}} \left[\Gamma_1 + (1-t)^{1+\frac{c}{2}}\Gamma_2 \right], &
for  $\epsilon= -1$;}
\label{matching1}
\end{eqnarray}
where
\begin{eqnarray}
\Gamma_1 &=& \frac{\Gamma(c) \Gamma(c-a-b) \Gamma(a+b-c+1) \Gamma(1-c)}
{\Gamma(c-a) \Gamma(c-b)\Gamma(b-c+1)\Gamma(a-c+1)} \\  \Gamma_2 &=&
\frac{\Gamma(c)\Gamma(a+b-c)\Gamma(c-a-b+1)\Gamma(1-c)}
{\Gamma(a)\Gamma(b)\Gamma(1-b)\Gamma(1-a)}
\end{eqnarray}
and
\begin{eqnarray}
\phi^+(r) + e^{i\omega_0}\phi^-(r) \to \cases{D t^{\frac{c
-2}{2}}(1-t)^{\frac{b  + a -c}{2}} \left[\bar\Gamma_1 +
(1-t)^{c-a-b}\bar\Gamma_2 \right],  & for  $\epsilon=1$; \cr D t^{\frac{c
-2}{2}} \left[\bar\Gamma_1 + (1-t)^{1+\frac{c}{2}}\bar\Gamma_2 \right], & for
$\epsilon= -1$;}
\label{matching2}
\end{eqnarray}
where
\begin{eqnarray}
\textstyle{\bar\Gamma_1 =
 \frac{\Gamma(c)\Gamma(c-a^+-b^+)\Gamma(a^++b^+-c+1)\Gamma(1-c)}{\Gamma(c-a^+)\Gamma(c-b^+)\Gamma(b^+-c+1)\Gamma(a^+-c+1)}
 + e^{i\omega_0}\frac{\Gamma(c)\Gamma(c-a^- -b^-)\Gamma(a^-+b^-
 -c+1)\Gamma(1-c)}{\Gamma(c-a^-)\Gamma(c-b^-)\Gamma(b^- -c+1)\Gamma(a^-
 -c+1)}}, \\ \textstyle{\bar\Gamma_2 = \frac{\Gamma(c)\Gamma(a^+ +b^+
 -c)\Gamma(c-a^+ -b^+
 +1)\Gamma(1-c)}{\Gamma(a^+)\Gamma(b^+)\Gamma(1-b^+)\Gamma(1-a^+ )} +
 e^{i\omega_0}\frac{\Gamma(c)\Gamma(a^- +b^- -c)\Gamma(c-a^- -b^-
 +1)\Gamma(1-c)}{\Gamma(a^-)\Gamma(b^-)\Gamma(1-b^-)\Gamma(1-a^-)}}
\label{}
\end{eqnarray}
Now comparing the respective coefficients  in Eq. (\ref{matching1}) and
Eq. (\ref{matching2}) we get the eigenvalue equation,
\begin{eqnarray}
f(E)=
    \frac{\Gamma(a)\Gamma(b)\Gamma(1-b)\Gamma(1-a)}{\Gamma(c-a)\Gamma(c-b)\Gamma(b-c+1)\Gamma(a-c+1)}=
    \frac{\sin(c-b)\pi\sin(c-a)\pi}{\sin a\pi \sin b\pi} = \mathcal
    M\frac{\cos(\beta +\omega_0/2)}{\cos(\alpha +\omega_0/2) },
\label{compare}
\end{eqnarray}
where
\begin{eqnarray}
\Gamma(a^\pm) = \chi_1 e^{\pm i\alpha_1},~\Gamma(b^\pm) = \chi_2 e^{\pm
i\alpha_2},~ \Gamma(1-a^\pm) = \chi_3 e^{\pm i\alpha_3},~ \Gamma(1-b^\pm) =
\chi_4 e^{\pm i\alpha_4},
\label{}
\end{eqnarray}
\begin{eqnarray}
\Gamma(c -a^\pm) = \lambda_1 e^{\pm i\beta_1},~\Gamma(c -b^\pm) = \lambda_2
e^{\pm i\beta_2},~ \Gamma(b^\pm -c +1) = \lambda_3 e^{\pm i\beta_3},~
\Gamma(a^\pm -c +1) = \lambda_4 e^{\pm i\beta_4},
\label{}
\end{eqnarray}
\begin{eqnarray}
\mathcal M =
\frac{\lambda_1\lambda_2\lambda_3\lambda_4}{\chi_1\chi_2\chi_3\chi_4},~ \beta
=\beta_1+\beta_2+\beta_3+\beta_4,~ \alpha =
\alpha_1+\alpha_2+\alpha_3+\alpha_4\,.
\label{}
\end{eqnarray}
The eigenvalue for general value of $\omega_0$ can be calculated
numerically. But we can immediately calculate the eigenvalue analytically at
least for some values of the extension parameter $\omega_0$ in the boundary
condition. So to appreciate constructing generalized boundary condition we now
investigate  some special cases.
\subsection{Case 1}
When the right hand side of Eq. (\ref{compare}) is infinity, we get $a = \pm
n$ or $b = \pm n$. $a= -n$ leads  to the eigenvalue, already calculated in
Ref. \cite{bellucci},
\begin{eqnarray}
E_{n,\,j,\,s} = \cases {\frac{\hbar^2}{2r^2_0} \left[\left(2n+ j_1+ \delta +1
\right)^2 - 4 -\frac{\omega^2 r_0^4}{\hbar^2}\right],& for $\epsilon=1$. \cr
-\frac{\hbar^2}{2r^2_0} \left[\left(2n+ j_1 -\delta +1 \right)^2 - 4
-\frac{\omega^2 r_0^4}{\hbar^2}\right]\,, & for $\epsilon= -1$.}
\label{spectrum}
\end{eqnarray}
The radial quantum number is given by
\begin{eqnarray}
n= \cases{0,1,\dots ,\infty, & for $\epsilon=1$.\cr  0,1,\dots ,n^{\rm
max}=[\delta/2-j-1], & for $\epsilon=-1$.}
\label{quantumno}
\end{eqnarray}
For $a = +n$ the energy spectrum will be the same expression (\ref{spectrum}),
with $n$ replaced by $-n$.  For $b= +n$, the energy spectrum will be
\begin{eqnarray}
E_{n,\,j,\,s} = \cases{\frac{\hbar^2}{2r^2_0} \left[\left(2n - j_1 -1 -\delta
\right)^2 - 4 -\frac{\omega^2 r_0^4}{\hbar^2}\right], & for $\epsilon=1$.\cr
-\frac{\hbar^2}{2r^2_0} \left[\left(-2n + j_1 +1 + \delta \right)^2 - 4
-\frac{\omega^2 r_0^4}{\hbar^2}\right], & for $\epsilon= -1$.}
\label{spectrum3}
\end{eqnarray}
for $b=- n$, $n$ in (\ref{spectrum3}) will be replaced by $-n$ and radial
quantum number $n$ is given in (\ref{quantumno}).
\subsection{Case 2}
We can also make the right hand side of Eq. (\ref{compare}) zero, which gives
us $c-b= \pm n$ or $c-a =\pm n$. for $c-b= +n$, the energy spectrum becomes,
\begin{eqnarray}
E_{n,\,j,\,s} = \cases{\frac{\hbar^2}{2r^2_0} \left[\left(-2n + j_1 + 1
-\delta \right)^2 - 4 -\frac{\omega^2 r_0^4}{\hbar^2}\right], & for
$\epsilon=1$.\cr -\frac{\hbar^2}{2r^2_0} \left[\left(2n - j_1 - 1 + \delta
\right)^2 - 4 -\frac{\omega^2 r_0^4}{\hbar^2}\right], & for $\epsilon= -1$.}
\label{spectrum4}
\end{eqnarray}
for $c-b=- n$, $n$ in (\ref{spectrum4}) will be replaced by $-n$ and radial
quantum number $n$ is given in (\ref{quantumno}). For $c-a =n$,
\begin{eqnarray}
E_{n,\,j,\,s} = \cases {\frac{\hbar^2}{2r^2_0} \left[\left(2n- j_1 -1 + \delta
\right)^2 - 4 -\frac{\omega^2 r_0^4}{\hbar^2}\right],& for $\epsilon=1$. \cr
-\frac{\hbar^2}{2r^2_0} \left[\left(2n- j_1 -1 -\delta  \right)^2 - 4
-\frac{\omega^2 r_0^4}{\hbar^2}\right]\,, & for $\epsilon= -1$.}
\label{spectrum5}
\end{eqnarray}
For $c-a = -n$, $n$ in (\ref{spectrum5}) will be replaced by $-n$ and radial
quantum number $n$ is given in (\ref{quantumno}).

\subsection{Case 3}
On the other hand if we make the right hand side $\pm 1$, then we get
degenerate(degenerate with respect to orbital quantum no $j_1$) eigenvalue.
For $c-b= +n+b$ and $c-a= +n+a$, we get,
\begin{eqnarray}
E_{n,\,s} =  \frac{\hbar^2}{2r^2_0} \left[\left(n+ \delta  \right)^2 - 4
-\frac{\omega^2 r_0^4}{\hbar^2}\right], \mbox{for} ~~\epsilon=1.
\label{spectrum6}
\end{eqnarray}
For $c-b=+n +b$ and $c-a= -n+a$ we get,
\begin{eqnarray}
E_{n,\,s} = -\frac{\hbar^2}{2r^2_0} \left[\left(n + \delta  \right)^2 - 4
-\frac{\omega^2 r_0^4}{\hbar^2}\right], \mbox{for} ~~\epsilon= -1.
\label{spectrum7}
\end{eqnarray}
%
\subsection{Case 4}

Even if, we can get totally degenerate eigenvalue when  $c-b = c-a \pm n$  and
the form  of the spectrum is given by

\begin{eqnarray}
E_{n } =\frac{\hbar^2}{2r^2_0}\left[ n^2 - 4- \frac{\omega^2
r_0^4}{\hbar^2}\right], \mbox{for}~~~\epsilon = +1\,.
\label{spectrum2}
\end{eqnarray}
For  $a+b+c=\pm n$ we get,
\begin{eqnarray}
E_{n } =-\frac{\hbar^2}{2r^2_0}\left[ n^2 - 4- \frac{\omega^2
r_0^4}{\hbar^2}\right], \mbox{for}~~~\epsilon = -1\,.
\label{spectrum2}
\end{eqnarray}
We have so far discussed the oscillator, where the dimension of the complex
coordinate is $N=2$. But we can generalize it for arbitrary dimensions $N > 1$. The
arbitrary dimensional conic oscillator Hamiltonian can be constructed from
conic oscillator of Ref. \cite{stefan} by making the magnetic field zero.
Once the oscillator Hamiltonian is given for general dimensions the rest of
the work of making self-adjoint extension is exactly same as what we have done
above.
\section{\small{\bf {Discussion}}} \label{con} 
In conclusion, we have calculated a generalized boundary condition for the
harmonic oscillator \cite{bellucci} and we have shown that this generalized
boundary condition can  restore the angular momentum degeneracy in energy
spectrum for a fixed value of the extension parameter. we have also recovered
the result of Ref.  \cite{bellucci} in our work. Not only that, we have shown
that it allows  more solutions for different values of the extension parameter.
\subsubsection*{Acknowledgements}
We  thank Palash B. Pal for comments on manuscript and helpful discussions.

\end{document}